\documentclass[a4paper,english,12pt,amssymb,nofootinbib,superscriptaddress]{revtex4-2}
\usepackage{lmodern}
\usepackage{lmodern}
\usepackage{xcolor}

\usepackage[T1]{fontenc}
\usepackage[latin9,utf8]{inputenc}
\usepackage{babel}
\usepackage{mathrsfs}
\usepackage{amsmath}
\usepackage{amssymb}
\usepackage{esint}
\usepackage{dsfont}
\usepackage[unicode=true,pdfusetitle,
bookmarks=true,bookmarksnumbered=false,bookmarksopen=false,
breaklinks=false,pdfborder={0 0 1},backref=false,colorlinks=false]
{hyperref}
\usepackage{xcolor}

\makeatletter

\@ifundefined{textcolor}{}
{%
	\definecolor{BLACK}{gray}{0}
	\definecolor{WHITE}{gray}{1}
	\definecolor{RED}{rgb}{1,0,0}
	\definecolor{GREEN}{rgb}{0,1,0}
	\definecolor{BLUE}{rgb}{0,0,1}
	\definecolor{CYAN}{cmyk}{1,0,0,0}
	\definecolor{MAGENTA}{cmyk}{0,1,0,0}
	\definecolor{YELLOW}{cmyk}{0,0,1,0}
}

\usepackage{babel}
\usepackage{babel}
\usepackage{babel}
\usepackage{babel}
\usepackage{babel}
\usepackage{babel}
\usepackage{babel}
\usepackage{babel}
\usepackage{babel}
\usepackage{babel}
\usepackage{babel}
\usepackage{babel}
\usepackage{babel}
\usepackage{babel}
\usepackage{babel}
\usepackage{babel}
\usepackage{babel}
\usepackage{graphicx}
\def\b{\begin{equation}}
\def\e{\end{equation}}

\@ifundefined{textcolor}{}{%
	\definecolor{BLACK}{gray}{0}
	\definecolor{WHITE}{gray}{1}
	\definecolor{RED}{rgb}{1,0,0}
	\definecolor{GREEN}{rgb}{0,1,0}
	\definecolor{BLUE}{rgb}{0,0,1}
	\definecolor{CYAN}{cmyk}{1,0,0,0}
	\definecolor{MAGENTA}{cmyk}{0,1,0,0}
	\definecolor{YELLOW}{cmyk}{0,0,1,0}
}

\usepackage{latexsym}\usepackage{bm}

\makeatother

\begin{document}
		\title{	Instability and Information Production Around Kerr Black Holes: Effects on Entropy and the Shadow}
	
	\author{Aydin Tavlayan}
	 
	\email{aydint@metu.edu.tr}
	
	\selectlanguage{english}%
	
	\affiliation{Department of Physics,\\
		Middle East Technical University, 06800 Ankara, Turkey}
	\author{Bayram Tekin}
	
	\email{bayram.tekin@bilkent.edu.tr}
	\affiliation{Department of Physics,\\
		Bilkent University, 06800 Ankara, Turkey}

	\selectlanguage{english}%
\begin{abstract}
\noindent  Massless or massive particles in unstable orbits around a Kerr black hole exhibit exponentially unstable motion when perturbed. They either plunge into the black hole or escape to infinity after making some oscillations around the equatorial plane. This exponentially unstable motion causes information production. In the case of the photons that escape to infinity, it was recently suggested that this information can be used to resolve the subring structure of the shadow image and obtain more precise data about the black hole mass and spin \cite{Strominger}. Here, we extend this method to obtain more precise results by including the {\it{non-equatorial}} contributions to the Lyapunov exponents. For massive particles plunging into the Kerr black hole, we show that the associated Kolmogorov-Sinai entropy derived from the Lyapunov exponents can be interpreted in the context of black hole thermodynamics and obeys Bekenstein's bound on the entropy of a physical system. Thus, the perturbed unstable orbits, either ending inside the black hole or at the observer's screen, have physical consequences. 
\end{abstract}
\maketitle

\section{Introduction}
The Event Horizon Telescope (EHT) collaboration published the images of the environments of two supermassive black holes, the one at the center of the M87 galaxy \cite{EHT1} and our own Sagittarius A* \cite{EHT2}. What makes these images possible at all is the existence of unstable null orbits around the black hole: light revolves around the black hole like a planet in certain orbits, but either plunges into the hole or escapes to infinity when slightly perturbed. So, a black hole with an environment is prone to observation. As noted in \cite{Strominger}, the images captured by the EHT Collaboration resolve the diameters of these black holes ($40\mu$as for M87 and $51.8\mu$as for Sgt A*) but not the {\it{subring structure}} that is produced by photons, which complete, say $n/2$ complete orbits before they escape to the telescope. It was shown in \cite{Strominger} that the first few subrings (produced by the $n=1$, $n=2$, and $n=3$ orbits) leave strong and universal signatures carrying information about the black hole parameters. These subrings are expected to be resolved by space-based sparse interferometric detectors. Here, one of our goals is to show that the addition of non-equatorial contributions from the accretion disc yields more precise information, especially for lower subring considerations. The contributions of the non-equatorial orbits were not considered in \cite{Strominger}.

In addition to the orbits of light, we also study the orbits of massive particles that are slightly perturbed and end up inside the black hole, and interpret the information content of these orbits and their relation to entropy. Possible connections between information theory and black hole thermodynamics were first proposed almost half a century ago \cite{Bekenstein0}. The subject started with the question of what happens if a spin-$1/2$ particle, of which every parameter is known except its spin, falls into the black hole. Unknown parameters introduce entropy according to information theory, and one should account for all possible states.  In this particular question, the spin is unknown, and because there are only two possibilities, the associated entropy contains an $\ln{2}$ term in units of the Boltzmann constant. We follow the same path here. We consider a particle with well-known position, mass, and energy orbiting in the vicinity of a Kerr black hole with given mass and rotation parameter. Then, we consider what happens if this particle is in an unstable orbit that is slightly perturbed, causing it to plunge into the black hole. In this case, we demonstrate that it will exhibit exponentially unstable motion, producing an enormous amount of information. We should note that all causal geodesics are Liouville-integrable around a Kerr black hole owing to the existence of four conserved quantities, including Carter's constant, energy, $z$ component of angular momentum, and the value of $g(p,p)$, where $g$ is the metric and $p$ is the four-momentum. Nevertheless, integrability does not require stability, and we are studying unstable orbits here. This increase in information can be associated with an increase in entropy, the Kolmogorov-Sinai entropy \cite{Dorfman}. It is crucial to state that this entropy is a mathematical concept based on the Lyapunov exponents of the exponentially unstable motion. Nevertheless, here we suggest that it can be interpreted as a physical entropy, and, as a check, we show that it respects the theory's bounds. This interpretation means that there is a significant increase in the total entropy of the particle-black hole system when the particle undergoes exponentially unstable motion and passes the event horizon. The bound we consider is the Bekenstein bound \cite{Bekenstein_bound}, which puts dimensional constraints on the physical system that is orbiting the black hole with a known energy and entropy, to protect the generalized second law of black hole thermodynamics when it plunges into the black hole \cite{Wald}. We assign the same amount of entropy that is produced by the exponentially unstable motion of the particle to a spherical volume by the method suggested in \cite{Visser} in Einstein's theory and extended to higher derivative gravity in \cite{Tavlayan}. We show that the entropy associated with this method satisfies the Bekenstein bound for a material system. This may be a new connection between information theory and black hole thermodynamics, and the concepts of this theory could be employed to explore statistical black hole thermodynamics, which may be a crucial step in finding a quantum gravity theory.

Let us note that the exponentially unstable motion around a black hole has been studied in the literature before: see \cite{Bombelli, Levin, Cornish, Dalui} and more recently \cite{Yunes}, whose conventions as well as the conventions of \cite{Strominger} and \cite{Lupsasca} have been followed here. The layout of the paper is as follows: In the next section, we discuss the basics of Lyapunov exponents and geodesic motion of the unstable orbits around the Kerr black hole. In Section III, we consider the case of massive orbits for which we calculate the associated Lyapunov exponents, the Kolmogorov-Sinai entropy, and its relation to the Bekenstein bound. In Section IV, we investigate the massless orbits that escape to infinity and demonstrate the connection between Lyapunov exponents and the subring structure of the shadow image. The current work appears to discuss two separate topics: entropy and shadow. However, from our vantage point, the unifying theme is the study of the physical and, perhaps, observable consequences of the unstable orbits of massless and massive particles around a Kerr black hole.    

\section{The Setup} 
\subsection{Lyapunov Exponents and Instability}
We start with the problem of instability for the particular case of geodesic motion around the Kerr black hole described by the metric in the $\left(t,r,\theta,\phi\right)$ coordinates with $\hbar=c=G=1$ units
\begin{eqnarray}
   ds^2 &=&-\left(1-\frac{2Mr}{\Sigma}\right)dt^2 - \frac{4Mar\sin^2\theta}{\Sigma}dtd\phi + \frac{\Sigma}{\Delta}dr^2 + \Sigma d\theta^2 \nonumber \\
   && +\left(r^2 + a^2 + \frac{2Ma^2 r\sin^2 \theta}{\Sigma}\right)\sin^2\theta d\phi^2,
\end{eqnarray}
where $a$ is the rotation parameter, and the metric functions read as
\begin{eqnarray}
    \Delta:=r^2 -2Mr+a^2, \hspace{2 cm} \Sigma:=r^2 + a^2 \cos^2\theta.
\end{eqnarray}
Using the two Killing symmetries $\xi_{(t)} = \partial_t$ and $\xi_{(\phi)}= \partial_\phi$ of the metric, we can reduce the phase space of the dynamics to the essential coordinates. Hence, the state vector of a particle or light can be defined as $\chi^{\mu}:=\left(r,\theta,p_r,p_{\theta}\right)$ of which Hamilton's equations govern the time evolution:
\begin{equation}
    \dot{\chi}^{\mu} = \Omega^{\mu\nu}\frac{\partial \mathcal{H}}{\partial \chi^{\nu}},\label{Hamilton1}
\end{equation}
where $\mathcal{H}$ represents the Hamiltonian and $\Omega^{\mu\nu}$ represents the symplectic matrix of the form
\begin{equation}
    \Omega^{\mu\nu}= \begin{pmatrix}
0_2 & \mathbb{I}_2\\
-\mathbb{I}_2 & 0_2 
\end{pmatrix},
\end{equation}
where $\mathbb{I}_2$ and $0_2$ represent the identity and zero matrices of rank 2, respectively. 
To quantify instability, let us assume that there are two trajectories in the phase space, denoted as $\chi^{\mu(0)}(t)$ and $\chi^{\mu}(t)$, infinitesimally close at the initial time $t=0$; and at a later time $t$, their separation can be defined as $\delta \chi^{\mu}(t):= \chi^{\mu}(t)-\chi^{\mu(0)}(t)$. The time-evolution of this separation follows from the linearization of Hamilton's equations (\ref{Hamilton1})
\begin{eqnarray}
    \delta \dot{\chi}^{\mu}(t)= J^{\mu}\,_{\nu}(t) \delta \chi^{\nu}(t),\hskip 1 cm \text{where} \hskip 1 cm 
    J^{\mu}\,_{\nu}(t):=\Omega^{\mu\rho}\partial_{\nu}\partial_{\rho} \mathcal{H}.
\end{eqnarray}
Here $J^{\mu}\,_{\nu}(t)$ is the evolution-Jacobian for the system. Direct integration yields
\begin{equation}
    \delta \chi^{\mu}(t) = e^{\int_{0}^{t} J^{\mu}\,_{\nu}(t')dt'} \delta \chi^{\nu}(0) =: L^{\mu}\,_{\nu}(t)\delta \chi^{\nu}(0),
\end{equation}
where in the second equality, "Lyapunov matrix" is defined 
\begin{equation}
    L^{\mu}\,_{\nu}(t):=e^{\int_{0}^{t} J^{\mu}\,_{\nu}(t')dt'}.
\end{equation}
The Lyapunov exponent is generically a complex number defined as a particular limit of this matrix
\begin{equation}\label{Lyapunov}
    \lambda :=\lim_{t\rightarrow \infty} \frac{1}{t} \log{\frac{L^{\mu}\,_{\mu}(t)}{L^{\nu}\,_{\nu}(0)}}.
\end{equation}
A positive $\lambda$ means that there is an instability in the system. Equivalently, the Lyapunov spectrum of the system can be analyzed by considering the eigenvalues of the evolution-Jacobian matrix $J^{\mu}\,_{\nu}(t)$, and the largest eigenvalue corresponds to the Lyapunov exponent. We shall use this approach in what follows. Next, we calculate the evolution matrix for the geodesics around the Kerr black hole.

\subsection{Hamiltonian of geodesic motion}
The Lagrangian of the massless and massive point particles can be taken to be 
\begin{equation}
    \mathcal{L}= \frac{1}{2} g_{\mu\nu}\frac{dx^{\mu}}{ds}\frac{dx^{\nu}}{ds},
\end{equation}
where $s$ is an affine parameter. The corresponding Hamiltonian is 
\begin{equation}
    \mathcal{H} = p_{\mu} \frac{dx^{\mu}}{ds}-\mathcal{L} =\frac{1}{2} g^{\mu\nu}p_{\mu}p_{\nu}=-m^2 ,
\end{equation}
where $m$ is the mass of the particle with $m \ge 0$. Explicitly, one has 
\begin{equation}
    \mathcal{H}= \frac{1}{2}\left(V + K\right) = -m^2,
\end{equation}
with the kinetic and potential parts given as 
\begin{equation}
    K := \frac{p_r^2}{g_{rr}} + \frac{p_{\theta}^2}{g_{\theta\theta}}, \hskip 2 cm  V :=\frac{L_z^2 g_{tt}+E^2 g_{\phi\phi}+2 E L_z g_{t\phi}}{g_{tt}g_{\phi\phi}-g_{t\phi}^2}, \label{potential}
\end{equation}
where $E=-p_t$ and $L_z=p_{\phi}$ are the conserved energy and angular momentum of the particle obtained from $\xi_{(t)}$ and $\xi_{(\phi)}$, respectively. With this Hamiltonian, one can calculate the corresponding evolution-Jacobian as
\begin{equation}\label{eqn14}
    J^{\mu}\,_{\nu}=\frac{1}{2}\begin{pmatrix}
\partial_r \left(\frac{p_r}{g_{rr}}\right) & \partial_{\theta} \left(\frac{p_r}{g_{rr}}\right) & \frac{1}{g_{rr}} & 0 \\
\partial_r \left(\frac{p_{\theta}}{g_{\theta\theta}}\right) & \partial_{\theta} \left(\frac{p_{\theta}}{g_{\theta\theta}}\right) & 0 & \frac{1}{g_{\theta\theta}} \\
-\partial_r \partial_r \left(K+V\right) & -\partial_{\theta} \partial_r \left(K+V\right) & -\partial_r \left(\frac{p_r}{g_{rr}}\right) & -\partial_r \left(\frac{p_{\theta}}{g_{\theta\theta}}\right)\\
-\partial_r \partial_{\theta} \left(K+V\right) & -\partial_{\theta} \partial_{\theta} \left(K+V\right) & -\partial_{\theta} \left(\frac{p_r}{g_{rr}}\right) & -\partial_{\theta} \left(\frac{p_{\theta}}{g_{\theta\theta}}\right)
\end{pmatrix}.
\end{equation}
The explicit form of this matrix is highly cumbersome to depict here. Consider the case for circular orbits, for which $p_r=0$:
\begin{equation}\label{eqn15}
    J^{\mu}\,_{\nu}=\frac{1}{2}\begin{pmatrix}
0 & 0 & \frac{1}{g_{rr}} & 0 \\
\partial_r \left(\frac{p_{\theta}}{g_{\theta\theta}}\right) & \partial_{\theta} \left(\frac{p_{\theta}}{g_{\theta\theta}}\right) & 0 & \frac{1}{g_{\theta\theta}} \\
-\partial_r \partial_r \left(K+V\right) & -\partial_{\theta} \partial_r \left(K+V\right) & 0 & -\partial_r \left(\frac{p_{\theta}}{g_{\theta\theta}}\right)\\
-\partial_r \partial_{\theta} \left(K+V\right) & -\partial_{\theta} \partial_{\theta} \left(K+V\right) & 0 & -\partial_{\theta} \left(\frac{p_{\theta}}{g_{\theta\theta}}\right)
\end{pmatrix}.
\end{equation}
If we further constrain this motion to the equatorial plane, $\theta=\frac{\pi}{2}$ and $p_{\theta}=0$, we can redefine the state vector as $\chi^{\mu}=\left(r,p_r\right)$ and the evolution-Jacobian becomes a $2\times 2$ matrix
\begin{equation}\label{Jacobian}
    J^{\mu}\,_{\nu}=\frac{1}{2}\begin{pmatrix}
0 & \frac{1}{g_{rr}}\\
-\partial_r \partial_r V & 0 
\end{pmatrix},
\end{equation}
which was studied in \cite{Yunes}. As noted above, the maximum eigenvalue of the evolution-Jacobian matrix is the Lyapunov exponent. Next, we study massive orbits.

\section{Massive Orbits}

Let us investigate massive particle orbits on the equatorial plane \cite{Frolov}. To do that, we need to focus on the radial part of the geodesic equation, which reads as
\begin{equation}
    \frac{dr}{ds}=\pm r^{-\frac{3}{2}}\sqrt{P},
\end{equation}
where
\begin{equation}
    P:=E^2 \left(r^3+a^2 r+2a^2 M\right)-4aMEL_z-\left(r-2M\right)L_z^2-rm^2\Delta.
\end{equation}
For the special case of the circular orbits, one can write the energy and angular momentum as a function of the independent parameter $r=r_{\text{circ}}$
\begin{equation}\label{EL}
    E_{{\text{circ}},\pm}=m\frac{r^2-2Mr\pm a\sqrt{Mr}}{r\sqrt{r^2-3Mr\pm 2a \sqrt{Mr}}}, \hskip 1 cm 
    L_{z,{\text{circ}},\pm}=m\frac{\pm \sqrt{Mr}\left(r^2 \mp 2a \sqrt{Mr} +a^2\right)}{r \sqrt{r^2-3Mr \pm 2a \sqrt{Mr}}},
\end{equation}
where $+$ sign corresponds to the prograde orbits, while $-$ sign corresponds to retrograde orbits. We study the stable and unstable orbits in the next two subsections as illustrative examples.
\subsection{Stable Orbit Case}
For the special case of the innermost {\it stable} circular orbits (ISCO), one can find the $r=r_{\text{ISCO}}$ as the solution of the polynomial 
\begin{equation}
    r^2-3a^2-6Mr \pm 8a \sqrt{Mr}=0.
\end{equation}
For the case with prograde innermost stable circular orbits with $M=1$, $m=1$, and $a=0.5$, one finds
\begin{eqnarray}
    &&r_{\text{ISCO},+}=4.233, \quad E_{\text{circ},+}=0.917882, \quad L_{z,\text{circ},+}=2.90287.
\end{eqnarray}
The corresponding Lyapunov exponents (\ref{Lyapunov}) turn out to be
\begin{equation}
    \lambda_{\text{pro},\pm}=0,
\end{equation}
as expected because these orbits are stable. (In what follows, we shall define entropy using the Lyapunov exponents; hence, stable orbits with vanishing exponents, such as the one studied above, will be assigned a zero entropy.)

\subsection{Unstable Orbit Case}
Among the equatorial circular orbits, for the special case of the binding orbits for which $E=m$, one can find the radius as
\begin{equation}
    r_{\text{bind},\pm}=2M\mp a+ 2 \sqrt{M\left(M\mp a\right)},
\end{equation}
which should be used to determine the particle's energy and angular momentum. One can calculate the potential and kinetic terms by using (\ref{potential}) as
\begin{eqnarray}
    &&V= \left.\frac{r \left(L_z^2-a^2 E^2\right)-2 M (L_z-a E)^2-E^2 r^3}{r \left(a^2+r (r-2 M)\right)}\right |_{r=r_{\text{bind}}}, \hskip 0.5 cm K=0.
\end{eqnarray}
The Lyapunov spectrum can be found as the eigenvalues of the evolution-Jacobian (\ref{Jacobian}), which are found for $r=r_{\text{bind}}$
\begin{eqnarray}\label{lambda1}
    \lambda_{\pm}=\pm \frac{1}{r^{5/2} \left(a^2+r (r-2 M)\right)}&&\left(2 a^4 M (L_z-a E)^2+r^3 \left(a^2 L_z^2-4 M^2 (L_z-a E) (9 L_z-7 a E)\right)\right.\nonumber\\
    &&\left.+6 M r^2 \left(a^2+4 M^2\right) (L_z-a E)^2-12 a^2 M^2 r (L_z-a E)^2\right.\nonumber\\
    &&\left.+6 M r^4 (a E-L_z) (a E-3 L_z)+2 E^2 M r^6-3 L_z^2 r^5\right)^{1/2}.
\end{eqnarray}
Even though we have an analytical formula, it is more illuminating to proceed numerically. However, one can analytically study various limits. For example, in the slowly rotating limit of the black hole, one has 
\begin{equation}
    \lambda_{\pm}=\lambda_{\pm}^\textbf{{Sch}} \left(1+a\frac{ (L_z-4 E M) (80 E M+9 L_z)}{4 \left(3 L_z^2 M-64 E^2 M^3\right)}\right) + \mathcal{O}\left(\left(\frac{a}{M}\right)^2\right),
\end{equation}
where the Schwarzschild black hole limit is given as 
\begin{equation}
\lambda_{\pm}^\textbf{{Sch}}=\pm \frac{\left(32 E^2 M^2- \frac{3L_z^2}{2}\right)^{1/2}}{16 M^2}.
\end{equation}

\begin{figure}
    \centering
    \includegraphics[width=0.75\linewidth]{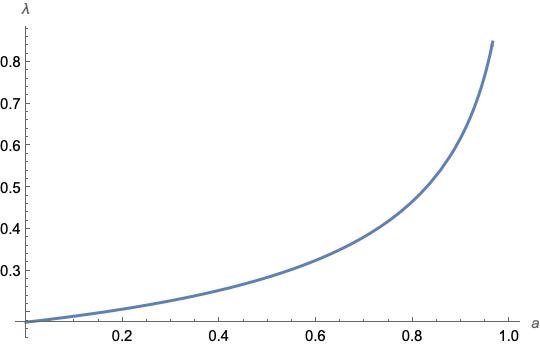}
    \caption{The Lyapunov exponent for the prograde binding orbit where the black hole parameter is  $M=1$ with $E_{\text{circ},+}=m=1$ is plotted as a function of the rotation parameter.}
    \label{Figpro}
\end{figure}

\begin{figure}
    \centering
    \includegraphics[width=0.75\linewidth]{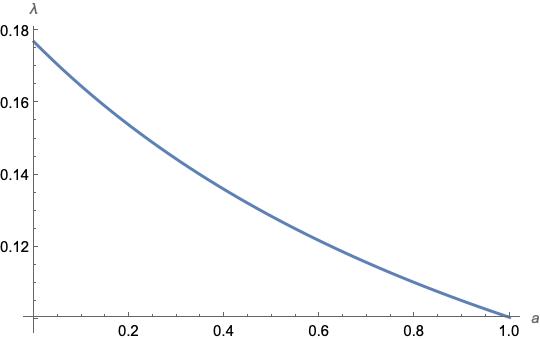}
    \caption{The Lyapunov exponent for the retrograde binding orbit where the black hole parameter is  $M=1$ with $E_{\text{circ},+}=m=1$ is plotted as a function of the rotation parameter.}
    \label{Figretro}
\end{figure}

\begin{figure}
    \centering
    \includegraphics[width=0.75\linewidth]{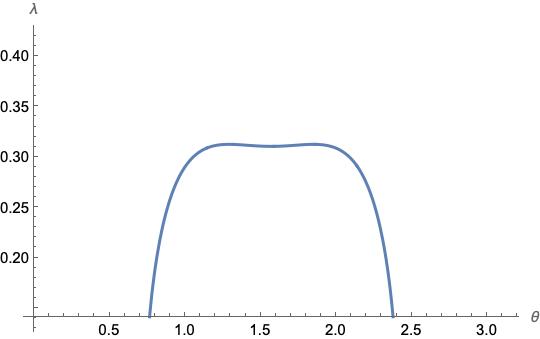}
    \caption{The angular dependence of the Lyapunov exponent for the case where $M=1$, $E=m=1$, $x=r=3$ and $\alpha=a=0.5$ is plotted.}
    \label{Figang}
\end{figure}

\begin{figure}
    \centering
    \includegraphics[width=0.75\linewidth]{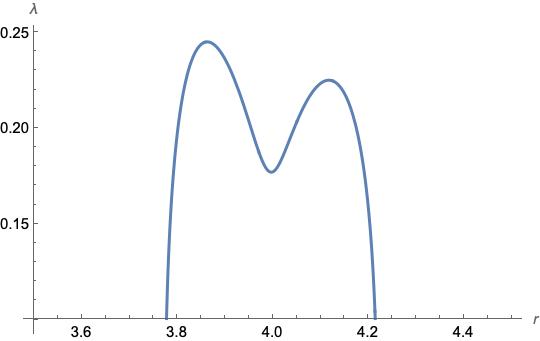}
    \caption{The radial dependence of the Lyapunov exponent for the case where $M=1$, $E=m=1$, $\theta=\pi/2$, and $\alpha=a=0.1$ is plotted.}
    \label{figa01}
\end{figure}

\begin{figure}
    \centering
    \includegraphics[width=0.75\linewidth]{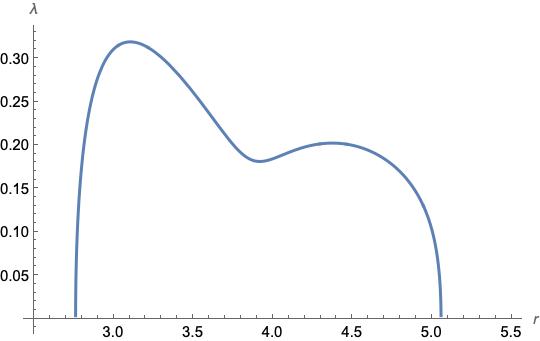}
    \caption{The radial dependence of the Lyapunov exponent for the case where $M=1$, $E=m=1$, $\theta=\pi/2$, and $\alpha=a=0.5$ is plotted.}
    \label{figa05}
\end{figure}

\begin{figure}
    \centering
    \includegraphics[width=0.75\linewidth]{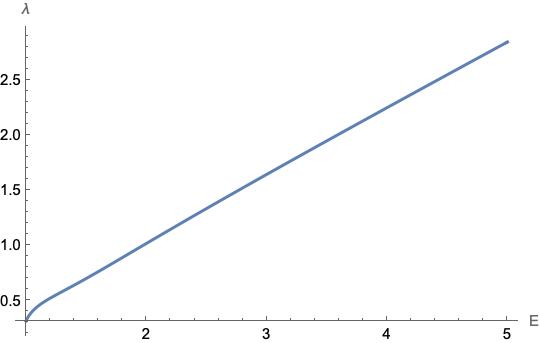}
    \caption{The plot shows the relation between the Lyapunov exponent and energy for the case where $M=1$, $m=1$, $x=r=3$, and $\alpha=a=0.5$, and the particle is on the equatorial plane when it is perturbed.}
    \label{fige}
\end{figure}

For this purpose, consider the prograde binding orbit:
\begin{equation}
    r_{\text{bind},+}=2M- a+ 2 \sqrt{M\left(M- a\right)},
\end{equation}
and assume the black hole parameters as  $M=1$ and $a=0.5$. These values yield
\begin{eqnarray}
    &&r_{\text{bind},+}=2.91421,\quad r_{\text{horizon}}=1.86603, \quad E_{\text{circ},+}=m=1, \quad L_{z,\text{circ},+}=3.41421.
\end{eqnarray}
As a result, from (\ref{lambda1}), the Lyapunov exponent can be computed to be 
\begin{equation}
    \lambda_{\text{pro},+}=0.284271.
\end{equation}
So, we can say that with a slight perturbation, a massive particle following the prograde binding orbit performs an exponentially unstable motion. One can see the relation between the Lyapunov exponent and the rotation parameter for the prograde binding orbit in the Fig. (\ref{Figpro}).
Similarly, for the retrograde binding orbit, we have
\begin{equation}
    r_{\text{bind},-}=2M+ a+ 2 \sqrt{M\left(M+ a\right)},
\end{equation}
for the same black hole parameters $M=1$ and $a=0.5$, one finds 
\begin{eqnarray}
    &&r_{\text{bind},-}=4.94949,\quad r_{\text{horizon}}=1.86603, \quad E_{\text{circ},-}=m=1, \quad L_{z,\text{circ},-}=-4.44949.
\end{eqnarray}
As a result, the positive Lyapunov exponent is found to be
\begin{equation}
    \lambda_{\text{retro},+}=0.128432,
\end{equation}
which shows that, for the retrograde binding orbit, perturbations lead to an exponentially unstable motion. One can see the relation between the Lyapunov exponent and the rotation parameter for the retrograde binding orbit in Fig. (\ref{Figretro}).
We stress that these results are for the equatorial-plane limits of the binding orbits. We can extend these results to the non-equatorial binding orbits for $2.91421 < r < 4.94949$. To do this, one can consider the radial part of the geodesic equation without the equatorial plane assumption. In this case, it reads as
\begin{eqnarray}
    &&\Sigma \frac{dr}{ds} = \pm \sqrt{\mathcal{R}(r)},
\end{eqnarray}
where
\begin{eqnarray}\label{R}
    &&\mathcal{R}(r):= \left(E\left(r^2+a^2\right)-a L_z\right)^2-\left(r^2-2Mr+a^2\right)\left(m^2 r^2 +\mathcal{Q}+\left(Ea-L_z\right)^2\right),
\end{eqnarray}
and $\mathcal{Q}$ is the Carter's constant related to a symmetric rank two Killing tensor. Explicitly, it reads
\begin{equation}
    \mathcal{Q}:=p_{\theta}^2+\cos^2{\theta}\left(\frac{L_z^2}{\sin^2{\theta}}+a^2\left(m^2-E^2\right)\right).
\end{equation}
Defining the dimensionless quantities, 
\begin{equation}
    \alpha:=\frac{a}{M},\hspace{0.5 cm} x:=\frac{r}{M}, \hspace{0.5 cm} \ell:=\frac{L_z}{Mm}, \hspace{0.5 cm} \tilde{Q}:=\frac{\mathcal{Q}}{M^2 m^2}, \hspace{0.5 cm} \tilde{E}:=\frac{E}{m},
\end{equation}
one can rewrite (\ref{R}) as
\begin{eqnarray}
    &&\mathcal{R}(x) := \left(\tilde{E}\left(x^2+\alpha^2\right)-\alpha \ell\right)^2-\left(x^2-2x+\alpha^2\right)\left(x^2 +\tilde{Q}+\left(\tilde{E}\alpha-\ell\right)^2\right).
\end{eqnarray}
Conditions for spherical orbits,
\begin{equation}
    \mathcal{R}(x)=0, \hspace{0.5 cm} \frac{d\mathcal{R}(x)}{dx}=0,
\end{equation}
can be solved for dimensionless angular momentum 
\begin{equation}
    \ell=-\frac{\alpha^3 \tilde{E}+\alpha^2 \sqrt{\alpha^2 x \left(\left(\tilde{E}^2-1\right) x+1\right)}+(x-2) x \sqrt{\alpha^2 x \left(\left(\tilde{E}^2-1\right) x+1\right)}-\alpha \tilde{E} x^2}{\alpha^2 (x-1)},
\end{equation}
and the dimensionless Carter's constant as
\begin{eqnarray}
    \tilde{Q}&=&\frac{x^2}{\alpha^3 (x-1)^2}\left(\alpha^3 \left(\left(2 \tilde{E}^2-1\right) x+1\right)+2 \alpha^3 \tilde{E} \sqrt{x \left(\left(\tilde{E}^2-1\right) x+1\right)}\right.\\
    &&\left.+\alpha x \left(x \left(-\left(\tilde{E}^2 ((x-4) x+5)\right)+(x-5) x+8\right)-4\right)+2 \alpha \tilde{E} (x-2) x \sqrt{x \left(\left(\tilde{E}^2-1\right) x+1\right)}\right).\nonumber
\end{eqnarray}
\begin{figure}
    \centering
    \includegraphics[width=0.75\linewidth]{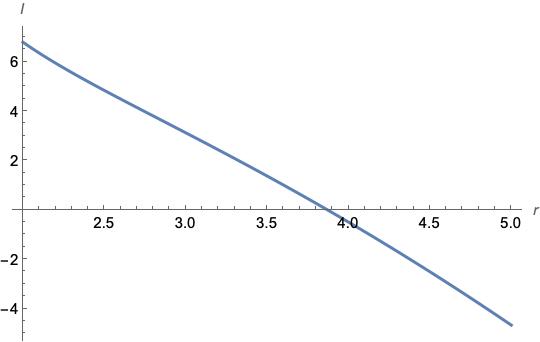}
    \caption{The angular momentum is plotted as a function of radius for the case where $M=1$, $m=E=1$, and $\alpha=a=0.5$.}
    \label{figanga05}
\end{figure}
\begin{figure}
    \centering
    \includegraphics[width=0.75\linewidth]{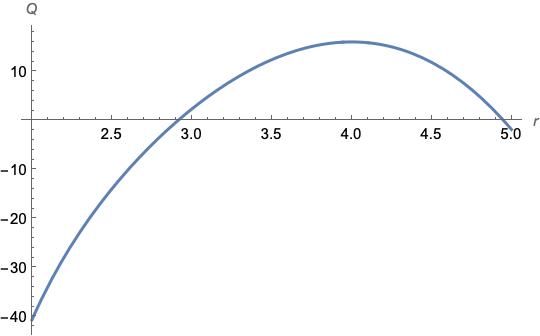}
    \caption{The Carter's constant is plotted as a function of radius for the case where $M=1$, $m=E=1$, and $\alpha=a=0.5$.}
    \label{figcartera05}
\end{figure}
For the special case of binding orbits, $E=m$ or $\tilde{E}=1$,
\begin{equation}
    \ell=\frac{\alpha^2-2 x^{3/2}+x^2}{\alpha-\alpha \sqrt{x}},
\end{equation}
and
\begin{equation}
    \tilde{Q}=\frac{x^2 \left(\alpha^2-\left(\sqrt{x}-2\right)^2 x\right)}{\alpha^2 \left(\sqrt{x}-1\right)^2}.
\end{equation}
To illustrate these relations, we plotted the angular momentum and Carter's constant as a function of radius for the case where $M=1$, $m=1$, and $\alpha=a=0.5$ in figures (\ref{figanga05}) and (\ref{figcartera05}), respectively. Since there is a non-zero Carter's constant, the motion is not bounded to the equatorial plane, which means $p_\theta \ne0$ and hence we must use (\ref{eqn14}) instead of (\ref{eqn15}) to calculate the Lyapunov exponents. As a result, instead of (\ref{lambda1}), we have to find the largest eigenvalue of the matrix (\ref{eqn14}) and use it as the Lyapunov exponent. The analytical formula is more complicated than (\ref{lambda1}), hence, we shall proceed numerically.
As an example, let us concentrate on the previous case where $M=1$ and $E=m=1$, so $x=r=3$ and $\alpha=a=0.5$. For this case, we have
\begin{eqnarray}
    &&\ell=3.12083,\quad \tilde{Q}=2.32497.
\end{eqnarray}
By assuming that the particle is on the equatorial plane when it is perturbed, the Lyapunov exponent can be computed to be
\begin{equation}
    \lambda_{\text{non-eq},+}=0.310321,
\end{equation}
which is larger than the Lyapunov exponents of both prograde and retrograde binding orbits. For non-equatorial spherical binding orbits, i.e., for particles which are not on the equatorial plane when perturbed, one can see the angular dependence of the Lyapunov exponent in Fig.(\ref{Figang}). As the figure shows, the instability first increases slightly for low orbital inclinations (around $\pi/2$, i.e., the equatorial plane), then decreases sharply for higher inclinations. Figures (\ref{figa01}) and (\ref{figa05}) show the radial dependence of the Lyapunov exponents for the binding orbits on the equatorial plane when they are perturbed, for different values of the rotation parameter. Figure (\ref{fige}) shows the relation between the Lyapunov exponent and the energy of the particle for the case where $M=1$, $m=1$, $x=r=3$, and $\alpha=a=0.5$, and the particle is on the equatorial plane when it is perturbed. It is crucial to note that for higher energy values, the Lyapunov exponent is positive, indicating instability.

\subsubsection{Normalization of the Lyapunov Exponents}
The Lyapunov exponent found via the evolution-Jacobian is not invariant under changes in the (time) parameterization. Hence, we need to find a relevant timescale to make it invariant. To achieve this, let us first study the angular part of the geodesic equation. The angular dependence is in radians.
\begin{eqnarray}
    &&\Sigma \frac{d\theta}{ds}=\pm \sqrt{\Theta(\theta)},
\end{eqnarray}
where
\begin{eqnarray}
    &&\Theta(\theta) := \mathcal{Q}-\cos^2{\theta}\left(a^2\left(m^2-E^2\right)+\frac{L_z^2}{\sin^2{\theta}}\right),
\end{eqnarray}
of which the non-dimensional form is
\begin{eqnarray}
    &&\tilde{\Theta}(\theta) := \tilde{Q}-\cos^2{\theta}\left(\alpha^2\left(1-\tilde{E}^2\right)+\frac{\ell^2}{\sin^2{\theta}}\right).
\end{eqnarray}

We first determine the angular turning points by solving $\tilde{\Theta}(\theta) = 0$. For simplicity, we define $u = \cos^2\theta$ and rewrite it as
\begin{equation}
    \tilde{\Theta}(u)=\tilde{Q}-u\left(\alpha^2\left(1-\tilde{E}^2\right)+\frac{\ell^2}{1-u}\right)=0,
\end{equation}
of which the solutions are
\begin{equation}
    u_{\pm}=\Delta_{\theta}\pm \sqrt{\Delta_{\theta}^2-\frac{\tilde{Q}}{\alpha^2\left(1-\tilde{E}^2\right)}}, \hskip 2 cm     \Delta_{\theta}=\frac{1}{2}\left(1+\frac{\tilde{Q}+\ell^2}{\alpha^2\left(1-\tilde{E}^2\right)}\right).
\end{equation}
Let us define the function \footnote{This integral can be expressed in terms of elliptic functions; we do not need the explicit forms here. For details, see \cite{Lupsasca}.}
\begin{equation}
    G_{\theta}:=\pm \int_{\theta_i}^{\theta_{f}} \frac{d\theta}{\sqrt{\Theta}},
\end{equation}
along the trajectory where $\theta_i$ and $\theta_f$ are the initial and final points. For one complete orbit, we can write
\begin{equation}
    G_{\theta}^1=2 \int_{\theta_{-}}^{\theta_{+}} \frac{d\theta}{\sqrt{\Theta}},
\end{equation}
where $\theta_-$ and $\theta_+$ are the angular turning points defined as $\theta_{\pm}:=\arccos{(\mp\sqrt{u_+})}$. As a result, the number of oscillations around the equatorial plane, along the trajectory, can be found as
\begin{equation}
    n=\frac{G_{\theta}}{G_{\theta}^1}.
\end{equation}
In section II, subsection A, we defined the Lyapunov exponent as a measure of temporal instability. To make this exponent an invariant quantity, we can redefine it with respect to the {\it{number}} of oscillations as
\begin{equation}\label{radialdev}
    \delta r(n) = e^{\lambda_P n} \delta r(0),
\end{equation}
where $\delta r(0)$ is the perturbation around the initial orbit. This new Lyapunov exponent, or the principal Lyapunov exponent, $\lambda_P$, is related to the old one as
\begin{equation}
    \lambda_P =\lambda_{+} \tau,
\end{equation}
where $\lambda_{+}$ is defined in (\ref{Lyapunov}) and $\tau = G_{\theta}^1$.
For the special case of the binding orbits, $E=m$ or $\tilde{E}=1$,
\begin{equation}
    \tilde{\Theta}(u)=\tilde{Q}-\frac{u \ell^2}{1-u}=0.
\end{equation}
Therefore,
\begin{equation}
    u=\frac{\tilde{Q}}{\tilde{Q}+\ell^2}.
\end{equation}

For the normalization of our exemplary case with $M=1$, $E=m=1$, $a=\alpha=0.5$, $r=x=3$, we can find
\begin{equation}
    G_{\theta}^1 := 2\int_{\theta_-}^{\theta_+}\frac{d\theta}{\sqrt{\tilde{\Theta}(\theta)}}=0.355709.
\end{equation}
In conclusion, the principal Lyapunov exponent becomes
\begin{equation}
    \lambda_P=\lambda_+ \tau=\lambda_+ G_{\theta}^1=0.114918.
\end{equation}
\subsection{Kolmogorov-Sinai entropy of the unstable orbits and the Bekenstein bound}
The Kolmogorov-Sinai (KS) entropy is a useful concept in quantifying the randomness of a deterministic dynamical system. Its full definition would constitute a long discussion for which we refer to Chapter 9 of the book \cite{Dorfman}. We only need Pesin's theorem here, which states that under certain conditions, the KS entropy, $ h_{KS}$, is equal to the sum of the positive Lyapunov exponents:
\begin{equation}
    h_{\text{KS}}=\sum_i \lambda_{+}^{i}. \label{KS}
\end{equation}
For the non-equatorial binding orbit with $M=1$, $E=m=1$, $r=x=3$ and $a=\alpha=0.5$, we
found that $\lambda_{\text{non-eq},P}=0.114918$. The corresponding 
Kolmogorov-Sinai entropy becomes
\begin{equation}
    h_{\text{KS}}=0.114918.
\end{equation}
The question is this: can we interpret this rather mathematical entropy from a physical point of view? If so, does this entropy satisfy the Bekenstein bound? To do this, we need to "resolve" the point particle in this unstable orbit, i.e., assign it a volume.  For this purpose, we shall use the method of \cite{Visser} in which one assigns an entropy to a bounded spherical volume~\footnote{Note that the robustness of this nice method was tested as it also works for some higher derivative gravity theories \cite{Tavlayan}.} with a process very similar to the one suggested by Gibbons and Hawking \cite{Gibbons}. Making proper adjustments in the action to introduce the volume constraint, one obtains the following saddle point approximation for the action,
\begin{equation}
    I_{\text{saddle}}=-\frac{1}{16 \pi G}\int d^Dx \sqrt{g}R=-\Omega_{D-2}\frac{R_V^{D-2}}{4G}=-\frac{A_V}{4G}.
\end{equation}
Using this, the spherical volume has the associated entropy 
\begin{equation}
    S=\frac{A_V}{4G}.
\end{equation}
The volume regularization and the entropy associated with a finite volume in spacetime require a complete discussion on its own, which is beyond the scope of this work. Here, we apply the technique found in \cite{Visser} to Einstein's gravity, and extended to higher curvature theories in \cite{Tavlayan}.  
At this point, let us identify the Kolmogorov-Sinai entropy ($h_{\text{KS}}=0.114918$) as the entropy of the bounded spherical volume. Then, its area, volume, and radius are
\begin{equation}
    A_V=0.459672, \hskip 1 cm V=0.0293053, \hskip 1cm R_V=0.191258.
\end{equation}
The Bekenstein bound is defined as
\begin{equation}
    S \le \frac{2 \pi E R}{\hbar}
\end{equation}
where we take $\hbar=1$, $R=R_V$ and $E=m=1$ for the binding orbit. This shows that
\begin{equation}
    S=0.114918 \le 2 \pi E R =1.20171.
\end{equation}
The Bekenstein bound is satisfied. One can conclude that interpreting the Kolmogorov-Sinai entropy as a physical entropy is viable, and this example lends support to Bekenstein's conjecture on the maximum entropy of material systems. Note that the discussion on Bekenstein's conjecture is still not finalized; see \cite{Wald} for a review of the problem.
We have finished our discussion of unstable massive orbits and the associated entropy. Next, we shall discuss the unstable light orbits, especially the connection between the Lyapunov exponents and the subring structure on the black hole shadow. 

\section{Massless Orbits}
The radial and angular parts of the geodesic equations \cite{Teo},\cite{Gralla} for a massless particle following an orbit that is not constrained to move on the equatorial plane are
\begin{eqnarray}
    &&\Sigma \frac{dr}{d\tilde{s}}=\sqrt{\mathcal{R}},\nonumber\\
    &&\Sigma \frac{d\theta}{d\tilde{s}}=\sqrt{\Theta},
\end{eqnarray}
where
\begin{eqnarray}
    &&\mathcal{R}=\left(r^2+a^2-a \ell\right)^2-\left(r^2-2 M r+a^2\right)\left(\tilde{Q}+\left(a-\ell\right)^2\right),\nonumber\\
    &&\Theta=\tilde{Q}+\cos^2{\theta}\left(a^2-\frac{\ell^2}{\sin^2{\theta}}\right),
\end{eqnarray}
and $\tilde{s}:=Es$, $\ell:=\frac{L_z}{E}$ and $\tilde{Q}:=\frac{\mathcal{Q}}{E^2}$. By using the spherical orbit conditions,
\begin{equation}
    \mathcal{R}=0, \hspace{0.5 cm} \frac{d\mathcal{R}}{dr}=0,
\end{equation}
one can write
\begin{eqnarray}
    &&\ell=-\frac{r^3-3Mr^2+a^2\left(r+M\right)}{a\left(r-M\right)},\nonumber\\
    &&\tilde{Q}=\frac{r^3}{a^2\left(r-M\right)^2}\left(4a^2M-r\left(r-3M\right)^2\right).
\end{eqnarray}
On the equatorial plane, two specific analytical expressions can be found, namely prograde and retrograde photon orbits, which are
\begin{equation}
    r_{\text{photon},\pm}=2M \left(1+\cos\left({\frac{2}{3}\arccos\left({\mp\frac{a}{M}}\right)}\right)\right).
\end{equation}
Between the prograde and retrograde orbits, one can find orbits restricted between $\theta_- \le \theta \le \theta_+ $ whose values can be found with the help of the angular part of the geodesic equation \cite{Himwich}. By defining $u:=\cos^2{\theta}$, one can solve
\begin{equation}
    \Theta=\tilde{Q}+u\left(a^2-\frac{\ell^2}{1-u}\right)=0
\end{equation}
to find the angular turning points, $\theta_{\pm}=\arccos{\left(\mp\sqrt{u_+}\right)}$, where
\begin{equation}
    u_{\pm}=\Delta_{\theta}\pm \sqrt{\Delta_{\theta}^2+\frac{\tilde{Q}}{a^2}}, \hspace{0.5 cm} \Delta_{\theta}=\frac{1}{2}\left(1-\frac{\tilde{Q}+\ell^2}{a^2}\right).
\end{equation}
As an example, consider a black hole with $a=0.5$ and $M=1$. For this case, we have 
\begin{equation}
    r_{+}=2.3473, \hspace{0.5 cm} r_{-}=3.53209,
\end{equation}
as limiting cases of the equatorial plane. Therefore, there is a non-equatorial orbit with $r=3$. For this orbit, we have
\begin{equation}
    \tilde{Q}=27, \hspace{0.5 cm} \ell=-1,
\end{equation}
so that is a retrograde orbit. For this orbit, the angular turning points become
\begin{equation}
    \theta_-=10.8462^{\circ}, \hspace{0.5 cm} \theta_+=169.154^{\circ}.
\end{equation}
Next, we compute the Lyapunov exponents for these non-equatorial light orbits and show their relation to the subring structure.
\subsection{Lyapunov Exponents and Shadow}
For this case, we do not restrict the motion to the equatorial plane. Henceforth, the evolution Jacobian becomes
\begin{equation}
    J^{\mu}_{\nu}=\frac{1}{2}\begin{pmatrix}
0 & 0 & \frac{1}{g_{rr}} & 0 \\
\partial_r \left(\frac{p_{\theta}}{g_{\theta\theta}}\right) & \partial_{\theta} \left(\frac{p_{\theta}}{g_{\theta\theta}}\right) & 0 & \frac{1}{g_{\theta\theta}} \\
-\partial_r \partial_r \left(K+V\right) & -\partial_{\theta} \partial_r \left(K+V\right) & 0 & -\partial_r \left(\frac{p_{\theta}}{g_{\theta\theta}}\right)\\
-\partial_r \partial_{\theta} \left(K+V\right) & -\partial_{\theta} \partial_{\theta} \left(K+V\right) & 0 & -\partial_{\theta} \left(\frac{p_{\theta}}{g_{\theta\theta}}\right)
\end{pmatrix}.
\end{equation}
The kinetic and potential terms are
\begin{equation}
    K = \frac{p_{\theta}^2}{g_{\theta\theta}}, 
\end{equation}
and
\begin{eqnarray}
    V&=&E^2\frac{\ell^2 g_{tt}(r,\theta)+g_{\phi\phi}(r,\theta)+2 \ell g_{t\phi}(r,\theta)}{g_{tt}(r,\theta)g_{\phi\phi}(r,\theta)-g_{t\phi}^2(r,\theta)}.
\end{eqnarray}
Since we are considering constant-radius spherical photon orbits, we took $p_r=0$.
The Lyapunov exponents are the eigenvalues of this matrix. The expressions are too cumbersome, so we did not write them here explicitly. For the specific case of $M=1$, $a=0.5$, and the orbit with radius $r=3$, let us assume that a slight perturbation occurs on the particle at the moment that the particle is at $\theta=\frac{\pi}{3}$. The corresponding angular momentum can be found as
\begin{equation}
    p_{\theta}=\sqrt{\tilde{Q}+\cos^2{\theta}\left(a^2-\frac{\ell^2}{\sin^2{\theta}}\right)}.
\end{equation}
The corresponding Lyapunov exponents become
\begin{eqnarray}
    &&\lambda_1=0-0.260123 i,\quad \lambda_2=0+0.260123 i,\quad \lambda_3=-0.597417,\quad \lambda_4=0.597417.
\end{eqnarray}
For normalization, we can use
\begin{equation}
    G_{\theta}^1 := 2\int_{\theta_-}^{\theta_+}\frac{d\theta}{\sqrt{\tilde{\Theta}(\theta)}}=1.16949.
\end{equation}
In conclusion, the principal Lyapunov exponent becomes
\begin{equation}
    \lambda_M=\lambda_+ \tau_M=\lambda_+ G_{\theta}^1=0.698674. \label{Lyapunovnoneq}
\end{equation}
If, on the other hand, this particle is perturbed while it is on the equatorial plane, $\theta=\frac{\pi}{2}$, we have
\begin{eqnarray}
    &&\lambda_1=0-0.148783 i,\quad \lambda_2=0+0.148783 i,\quad \lambda_3=-0.580475,\quad \lambda_4=0.580475.
\end{eqnarray}
For normalization, we can use
\begin{equation}
    G_{\theta}^1 := 2\int_{\theta_-}^{\theta_+}\frac{d\theta}{\sqrt{\tilde{\Theta}(\theta)}}=1.16949.
\end{equation}
In conclusion, the principal Lyapunov exponent becomes
\begin{equation}
    \lambda_M=\lambda_+ \tau_M=\lambda_+ G_{\theta}^1=0.678861. \label{Lyapunoveq}
\end{equation}
The most important consequence of these results is that, even though most particles from the accretion disc are assumed to originate from the equatorial plane, considering deviations from this plane is necessary for a more accurate result when relaxing this assumption. This is best seen by focusing on the black hole shadow. In our previous work, we investigated the black hole shadows in detail \cite{Tavlayan2}. Nevertheless, in that paper, we assumed that photons emitted from an orbiting black hole make an infinite number of turns around it. In other words, we did not consider the subring structure of the shadow image. Let us consider a photon hitting a point at radius $\rho=\rho_c + \delta\rho$  on the image plane where the shadow edge is at $\rho_c$. The radial deviation on the image plane, $\delta\rho$, is linearly related to the radial deviation around the black hole caused by perturbation, $\delta r(0)$ in \ref{radialdev}. As a result, for such a photon to orbit $\frac{n}{2}$ times before hitting the image, it must satisfy
\begin{equation}
    \frac{\delta\rho_n}{\rho_c}\approx e^{-\lambda_M n}
\end{equation}
around the shadow edge \cite{Strominger}. This is an approximate relation used in \cite{Strominger} as equation (11). One can understand this relation as follows: Subrings corresponding to fewer oscillations are visible and easier to interpret. Yet as the number of oscillations increases, they come closer together and approach the shadow limit, making them harder to resolve. Nonetheless, we are interested in a smaller number of subrings for observational and interferometric details, where small deviations in the Lyapunov exponent are important. Hence, the visible and consequential difference between the Lyapunov exponents considered on the equatorial  (\ref{Lyapunoveq}) and on the non-equatorial orbits (\ref{Lyapunovnoneq}) should be considered, which was one of the main aspects of the current paper.

\section{Conclusions}

We considered some exponentially unstable motion undergone by massless and massive particles in the vicinity of a Kerr black hole. This leads us to two significant consequences. First, we showed that for massive particle orbits, the exponentially unstable motion of the massive particles that end inside the black hole can be interpreted in terms of black hole thermodynamics via the relation between the Lyapunov exponents and the Kolmogorov-Sinai entropy. Even though the KS entropy is an a priori mathematical construct, here we suggested a physical interpretation that obeys the expected bounds of the theory, such as the Bekenstein bound \cite{Bekenstein_bound}. This interpretation required us to introduce a physical volume for the particle on the unstable massive orbit. For this purpose, we used the method of \cite{Visser}. The fact that the Kolmogorov-Sinai entropy can be interpreted as a physical entropy that obeys the entropy bound conjecture is a promising result that may help build a bridge between information theory and black hole thermodynamics. Second, we investigated the instability of light and extended the method used \cite{Strominger} to cover non-equatorial-plane contributions to the black hole image. This addition provides nontrivial contributions to the subring structure of the shadow image, enabling a precise determination of the black hole mass and angular momentum. In general, as a viable approximation, photons emanating from the accretion disc on the equatorial plane are considered when constructing the shadow image. This does not cause any problems for the higher-order subrings or the shadow itself, for which it is assumed ($n \rightarrow \infty$). Nevertheless, deviations caused by non-equatorial photons may be relevant for lower-order subrings, especially for $n=1$, $n=2$, and $n=3$, and one should consider them to obtain more accurate results for observable parameters. An outstanding problem we have not solved is to assign a Kolmogorov-Sinai type entropy for the null orbits: at this stage, we have not found a convincing volume regularization for these null orbits a la \cite{Visser}. The method discussed can also be applied to different theories of gravity and to observational tools such as gravitational waves and near-horizon studies. \cite{Deich2022},\cite{Cornish},\cite{Kan},\cite{Lyu},\cite{Hashimoto}.

\end{document}